# Second-Coordination-Sphere Cation Substitution as a Tool for Controlling Phase Transitions and Performance of the Luminescence Thermometry

Muhammad T. Abbas[1], M. Szymczak[1], M. Fandzloch[1], D. Szymanski[1], A. Sieradzki[2], L. Marciniak[1,*]

[1] Institute of Low Temperature and Structure Research, Polish Academy of Sciences, Okólna 2, 50-422 Wrocław, Poland

[2] Department of Experimental Physics, Wrocław University of Technology, Wybrzeże Wyspiańskiego 27, 50-370 Wrocław, Poland

*corresponding author: l.marciniak@intibs.pl

**Abstract**

Despite the exceptionally high relative sensitivities achieved by luminescent thermometers based on first-order structural phase transitions, their principal limitation lies in the inherently narrow thermal operating range associated with the transition temperature. In this work, we demonstrate that partial substitution of $Li^+$ by $Na^+$ ions in the second coordination sphere of $Eu^{3+}$ ions in $LiYO_2$ enables a substantial shift of the phase transition temperature, thereby allowing controlled optimization of the thermometric performance. This approach represents a significantly more cost-effective and efficient strategy for tuning the phase transition temperature compared with the previously proposed substitution of $Y^{3+}$ by other lanthanide ions. Importantly, we show that lowering the transition temperature through Na$^+$ incorporation simultaneously introduces static compositional disorder and local lattice strain. As a consequence, the enthalpy difference between the competing structural phases decreases, and the cooperativity of the lattice distortion is reduced, indicating a gradual weakening of the first-order character of the phase transition. Our results demonstrate that such structural modifications, while effective in shifting the transition temperature, inevitably lead to a reduction in the relative sensitivity of phase-transition-based luminescent thermometers.

**Introduction**

The spectroscopic properties of lanthanide ions (Ln$^{3+}$) are commonly regarded as only weakly sensitive to variations in their local environment[1–3]. This characteristic originates from the fact that the 4*f* orbitals, within which intraconfigurational electronic transitions responsible for lanthanide luminescence occur, are effectively shielded from the surrounding lattice by the outer 5*s* and 5*p* orbitals. As a consequence, the spectral positions of 4*f*-4*f* transitions of Ln$^{3+}$ ions vary only marginally among different host materials[3–5]. Although the energies of these transitions remain largely invariant, the shape of the emission bands can change substantially when the symmetry of the crystallographic site occupied by Ln$^{3+}$ ions is altered. This effect arises from changes in the Stark splitting of the 4*f* multiplets, manifested as variations in both the number of Stark components and the magnitude of their energetic separation[4]. Such symmetry-driven spectral modifications, particularly those induced by thermally activated structural phase transitions in certain host lattices, have been successfully exploited in the development of phase-transition-based luminescence thermometers[6–15]. As demonstrated in earlier studies, a change in the point symmetry of the crystallographic site occupied by Ln$^{3+}$ ions across a phase transition is accompanied not only by a pronounced reshaping of individual emission bands but also by modifications of the luminescence kinetics[6,10,16]. These kinetic changes arise from variations in the radiative depopulation probability of the emitting level. In such systems, the Stark components characteristic of the low-temperature (LT) phase progressively vanish as the phase transition temperature is approached, while those associated with the high-temperature (HT) phase emerge and increase in intensity [17–19]. The abrupt nature of this spectral redistribution gives rise to exceptionally high relative sensitivities of luminescence thermometers, reaching values as high as 30% K$^{-1}$ [7].

Despite this advantage, the primary limitation of phase-transition-based luminescence thermometers is their relatively narrow operational temperature window, which is typically confined to the immediate vicinity of the phase transition temperature[8,12]. One strategy proposed to overcome this limitation involves introducing additional dopant ions with ionic radii significantly different from that of the host cation[8,12,17]. The resulting lattice strain modifies the average metal-oxygen bond lengths and, consequently, alters the thermal energy required to induce the structural transition, leading to a shift in the phase transition temperature.

In luminescence thermometers based on Ln$^{3+}$ emission, this approach commonly employs other lanthanide ions, preferably optically inactive ones, to avoid activating additional energy-transfer pathways that could reduce emission intensity[8,12,17]. However, due to the lanthanide contraction, the maximum difference in ionic radii between La$^{3+}$ and heavier Ln$^{3+}$ ions is relatively small ($R_{La3+}$=1.16 pm and $R_{Lu3+}$=0.97pm) [20]. As a result, substantial shifts in the phase transition temperature typically require very high dopant concentrations. Such high doping levels are economically unfavourable due to the high cost of lanthanide precursors and may also introduce significant structural disorder.

In this work, we present an alternative strategy based on cation substitution within the second coordination sphere of Ln$^{3+}$ ions (Figure 1). As a model system, we employ LiYO$_2$, which undergoes a reversible first-order structural phase transition from a low-temperature monoclinic phase to a high-temperature tetragonal phase near 320 K, accompanied by a change in the point symmetry of the crystallographic site occupied by $Y^{3+}$ (and substituted by $Ln^{3+}$ ions) from $C_1$ to $D_{2d}$ [19,21–26]. Previous studies have shown that lowering the phase transition temperature to approximately 160 K requires substitution of up to 40% of $Y^{3+}$ ions with $Yb^{3+}$ [12]. Here, we demonstrate that a comparable shift of the phase transition temperature can be achieved by substituting only 15% of $Li^+$ ions with $Na^+$ ions, owing to the much larger difference in ionic radii between $Li^+$ and $Na^+$. This approach significantly reduces the cost of tuning the thermometric properties, as $Na^+$ precursors are considerably less expensive than lanthanide salts. Importantly, our results reveal that an equivalent effect on the phase transition temperature can be achieved by modifying the $Ln^{3+}$-$O^{2-}$ bond length through unit-cell expansion induced by $Na^+$ substitution, in analogy to the effect obtained by substituting $Y^{3+}$ with $Yb^{3+}$ ions. A well-known drawback of strategies aimed at shifting the operational temperature range of phase-transition-based luminescence thermometers through chemical substitution is the concomitant reduction in relative sensitivity. In this work, we provide, for the first time, a mechanistic explanation for this phenomenon, demonstrating that it arises from a doping-induced change in the nature of the structural phase transition itself.

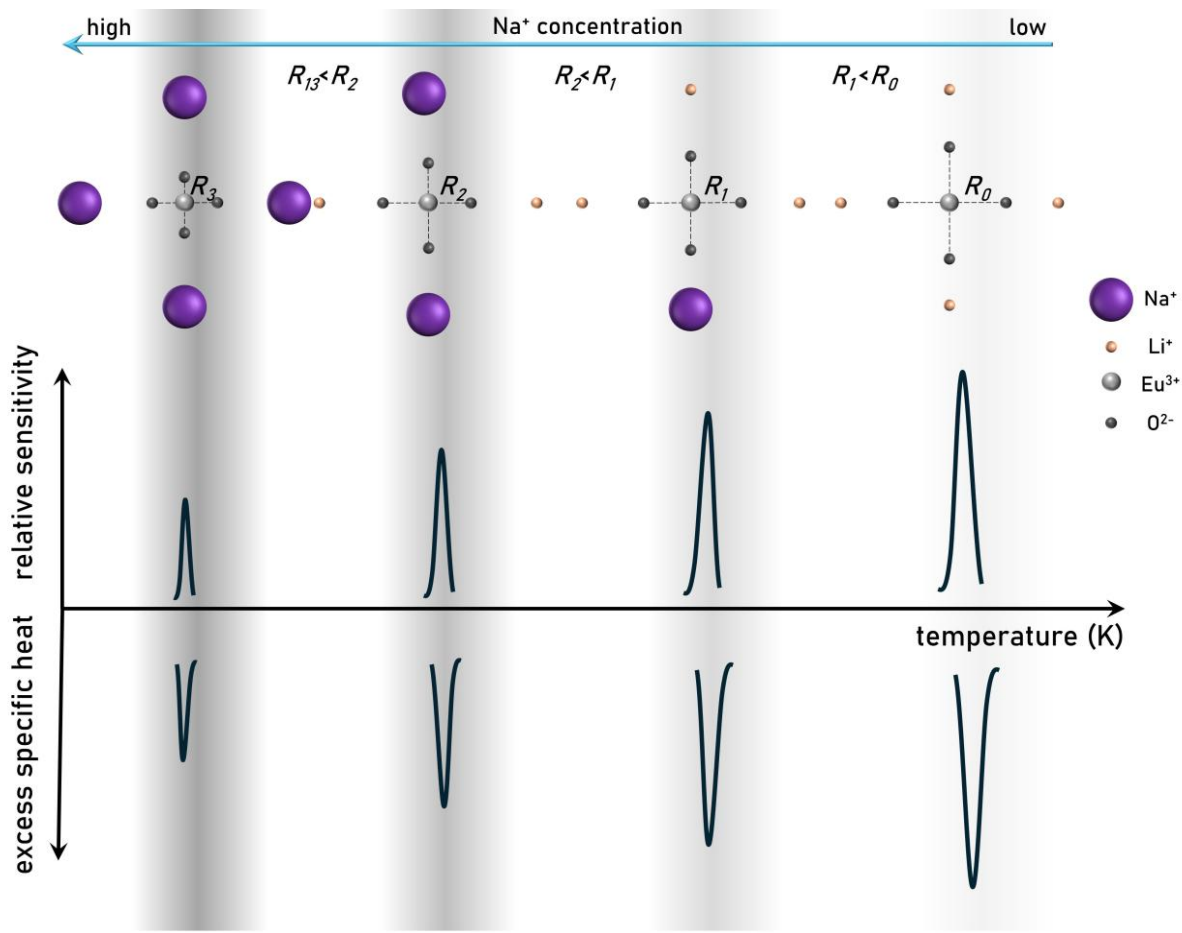

**Figure 1**. The main concept of the work: in the LiYO$_2$:Eu$^{3+}$ around 320 K the phase transition from LT to HT phases in observed. However, when the Na$^+$ ions replace the Li$^+$ ions of the host material a change in the phase transition temperature is observed. Namely an increase in the Na$^+$ ions concentration in the second coordination sphere of the Eu$^{3+}$ ions leads to the shortening of the Eu$^{3+}$-O$^{2-}$ bond length $R$ resulting in the reduction of the phase transition temperature. The structural modification induced by this doping results in the change of the character of the phase transition and the reduction in the excess specific heat causing the reduction in the relative sensitivity of the ratiometric luminescence thermometer based on the structural phase transition.

**Experimental Section**

*Materials and synthesis*

LiYO$_2$:1%Eu$^{3+}$, LiYO$_2$:1%Eu$^{3+}$, x%Na$^+$ (x = 0.5, 1, 2, 5, 10, 15, 20) samples were synthesized by the high-temperature solid state method. Li$_2$CO$_3$ (99.9% of purity, Chempur), Y$_2$O$_3$ (99.999% of purity, Stanford Materials Corporation), Na$_2$CO$_3$ (99.9% of purity, Alfa

Aesar), and $Eu_2O_3$ (99.99% of purity, Stanford Materials Corporation) were used as raw materials. The raw materials were weighed according to the stoichiometric ratios, put into an agate mortar, and ground with the addition of a few drops of hexane to achieve a homogeneous mixture. Then, the mixture was transferred to an alumina crucible and sintered at 1273 K for 6 hours with the heating rate of 10 K min$^{-1}$. Finally, the samples were naturally cooled to room temperature and ground again for further characterization.

*Methods*

The purity of synthesized samples was examined by using the powder X-ray diffraction technique. Powder diffraction measurements were performed in Bragg-Brentano geometry using a PANalytical X'Pert Pro diffractometer equipped with an Oxford Cryosystems Phenix low-temperature unit and an Anton Paar HTK 1200N high-temperature attachment using Ni-filtered Cu K$\alpha$ radiations ( V = 40 kV, I = 30 mA). Diffraction data were collected in a 10 - 90º 2θ range. ICSD database entries 50992 (LT phase) and 50993 (HT phase) were taken as initial models for analysis of the obtained diffraction data. The elemental composition of the powders was determined by ICP–OES using an iCAP™ 7400 spectrometer. Approximately 2–5 mg of each sample was digested in an $HNO_3$/HCl (1:1, 4 mL) mixture and diluted with distilled water (40 mL) prior to analysis. Measurements were performed in triplicate for Na, Li, Y, and Eu, using two analytical emission lines for the latter two elements. Reported concentrations represent average values with corresponding standard deviations.

Scanning electron microscopy (SEM) was used to investigate the sample morphology, and the elemental distribution was examined via energy-dispersive X-ray spectroscopy (EDS). These characterizations were performed using an FEI Nova NanoSEM 230 microscope equipped with an EDAX Genesis XM4 detector, operated at an accelerating voltage of 30 kV for SEM imaging and 5kV for EDS mapping. For sample preparation, the powders were dispersed in methanol, and a droplet of the suspension was deposited onto a carbon stub and dried prior to imaging.

Calorimetric measurements were performed using a Mettler Toledo DSC-3 differential scanning calorimeter. The samples were sealed in aluminum pans, and nitrogen was employed as the purge gas. Measurements were conducted at a constant heating rate of 5 K min$^{-1}$ over the temperature range of 120-350 K. The excess heat capacity associated with the phase transition was determined by subtracting an appropriate baseline from the measured heat-flow data.

Raman measurements were performed using an Edinburgh Instruments RMS-1000 Raman microscope, operating with a 457 nm excitation laser, a 20x objective lens, and an 1800 lines mm-1 diffraction grating. Temperature-dependent experiments were carried out under controlled conditions using a Linkam THMS600 heating-cooling stage.

The emission spectra under 395 nm excitation were collected using the FLS1000 Fluorescence Spectrometer from Edinburgh Instruments equipped with a 450W Xenon lamp and R928 photomultiplier tube from Hamamatsu. To perform the temperature-dependent measurements, the temperature of the sample was controlled using a THMS600 heating-cooling stage from Linkam (0.1 K temperature stability and 0.1 K point resolution).

**Results and discussion**

It is well established in the literature that LiYO$_2$ crystallizes in two distinct polymorphic forms: a monoclinic low-temperature (LT) phase (space group *P2$_1$/c*) and a tetragonal high-temperature (HT) phase (space group *I4$_1$/amd*)[19,21–25,27]. Temperature is the most extensively studied driving force for the phase transformation. Upon heating, LiYO$_2$ undergoes a reversible first-order structural phase transformation, resulting in a more ordered crystal lattice and higher symmetry in the HT phase. The differences between the LT and HT structures extend beyond the variation and reduced dispersion of the Y$^{3+}$-O$^{2-}$ bond lengths observed in the HT phase and involve a pronounced reorganization of the local coordination environment of lithium (Figures 2a and 2b). In the LT phase, each $Y^{3+}$ ion is linked through oxygen atoms to seven $Li^+$ ions, with lithium adopting a three-fold oxygen coordination. Upon transition to the HT phase, the local environment changes substantially: eight $Li^+$ ions become connected to $Y^{3+}$ via oxygen bridges, while the coordination number of $Li^+$ increases from three to four. This extensive rearrangement of the local structure has been demonstrated to exert a strong influence on the spectroscopic properties of $LiYO_2$ doped with $Ln^{3+}$ ions. Previous studies have almost exclusively focused on structural modifications within the yttrium sublattice. Substitution of $Y^{3+}$ with luminescent $Ln^{3+}$ activators, as well as co-doping with optically inactive ions, has been employed to systematically adjust metal-oxygen bond lengths, thereby tuning the phase transition temperature and the associated luminescent response of the $Ln^{3+}$ centers. In contrast, no reports have addressed structural engineering strategies that target coordination environments other than yttrium, in particular the lithium sublattice, as a means of modulating $Ln^{3+}$ luminescence. Accordingly, in this study $Li^+$ was partially substituted with $Na^+$, while maintaining a constant $Eu^{3+}$ concentration of 1%. X-ray diffraction analysis confirms that $Na^+$ incorporation up to 15% preserves phase purity, with no evidence of secondary phases. At higher $Na^+$ contents (20%), additional diffraction reflections emerge, indicating the onset of impurity phase formation (Figure S1). Accordingly, only samples containing up to 15% $Na^+$ are considered in the subsequent analysis. The ICP analysis confirmed the well agreement between the intentional $Na^+$ concentration and real concentration of dopant ions (Figure S2). SEM analysis coupled with EDS mapping for a representative $LiYO_2$:1%$Eu^{3+}$,1%$Na^+$ sample revealed that high-temperature synthesis leads to the formation of microparticles with irregular and non-uniform morphology, which are typical for this preparation method. The uniform spatial distribution of Li, Y, O, Eu, and Na across the analyzed areas (Figure 2d) clearly demonstrates the effective incorporation of dopants into the LiYO$_2$ matrix without detectable phase segregation or compositional inhomogeneities.

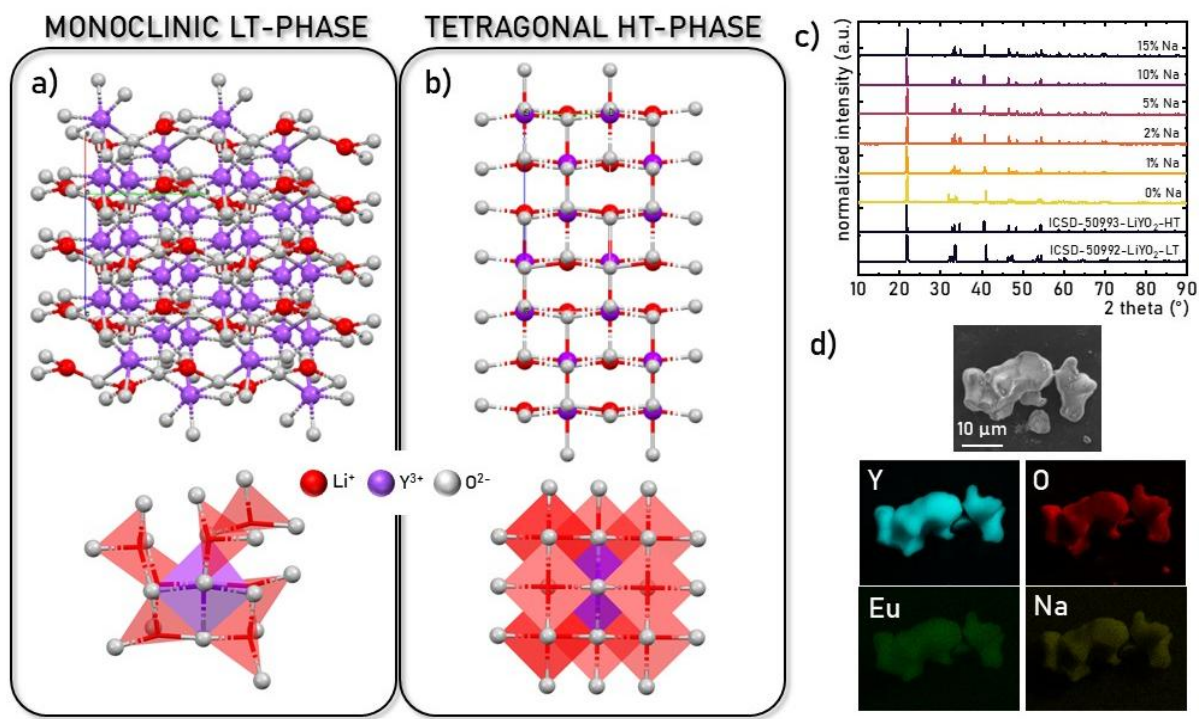

**Figure 2.** Visualization of the monoclinic LT-phase -a) and the tetragonal HT-phase -b) of LiYO$_2$, with structural fragments illustrating the coordination environment of Y$^{3+}$ ions, including Li$^+$ ions coordinated by oxygens. X-ray diffraction patterns of LiYO$_2$:1%Eu$^{3+}$ samples with varying Na$^+$ dopant concentrations -c). Representative SEM image together with EDS elemental maps (Li, Y, O, Eu, Na) for LiYO$_2$:1%Eu$^{3+}$, 1%Na$^+$ -d).

To verify the structural changes associated with the temperature-induced phase transition, temperature-dependent Raman spectroscopy measurements were performed for all investigated samples. Representative spectra for the 1%Na$^+$ and 1%Eu$^{3+}$ doped LiYO$_2$ sample are presented in Figure 3a, whereas the remaining datasets are provided in the Supporting Information (Figure S3). For the 1%Na$^+$ sample, a distinct phase transition is observed in the temperature range of approximately 303-308 K. This transition is manifested by pronounced changes in both the position and intensity of Raman bands. The most significant modifications occur in the spectral region around 550 cm$^{-1}$, which corresponds to the internal vibrational modes of YO$_6$ octahedra, namely O-Y-O bending and Y-O stretching vibrations. A general reduction in the number of Raman-active modes within the 450-600 cm$^{-1}$ range is observed upon heating, indicating a transition toward a structure of higher crystallographic symmetry. This behavior is consistent with the monoclinic-to-tetragonal transformation of LiYO$_2$, where the tetragonal phase exhibits a narrower distribution of Y-O bond lengths and reduced distortion of the YO$_6$ octahedra compared to the monoclinic structure. Consequently, the vibrational mode splitting decreases, leading to fewer observable Raman bands. Based on the temperature evolution of the Raman spectra, the fraction of the low-temperature phase of LiYO$_2$ was determined for Na$^+$ concentrations ranging from 0% to 5% (Figure 3b). As the Na$^+$ content increases, the phase transition temperature systematically shifts toward lower values, indicating that Na$^+$ incorporation stabilizes the tetragonal phase. Furthermore, for the 5%Na$^+$ sample, the transition becomes noticeably less abrupt, suggesting a gradual change in the character of the phase transition from first-order toward second-order behavior. This observation will be further discussed in light of the DSC results. The reduced sharpness of the transition is also clearly reflected in Figure 3c, which compares luminescence maps for the 1%Na$^+$ and 15%Na$^+$ samples. Differential scanning calorimetry (DSC) was used to examine the effect of Na$^+$ substitution at the Li$^+$ site on the first-order phase transition in LiYO$_2$:Eu$^{3+}$. Figure 3d displays the excess specific heat curves recorded upon heating and cooling for representative Na$^+$ concentrations (Figure S4). The undoped compound exhibits a sharp and intense calorimetric anomaly, characteristic of a well-defined first-order phase transition accompanied by a pronounced thermal hysteresis. Upon Na substitution, the excess specific heat anomalies systematically shift to lower temperatures, as summarized in Figure 3e, which plots the transition temperatures extracted from the peak positions as a function of Na$^+$ content. Even a small amount of Na$^+$ substitution results in a substantial suppression of the transition temperature, while further Na$^+$ incorporation leads to a continuous and monotonic downward shift of both heating and cooling transition temperatures. In parallel, the calorimetric anomalies become progressively broader and weaker with increasing Na$^+$ concentration (Figure 3d). The gradual reduction of the excess specific heat peak area indicates a marked decrease of the latent heat associated with the transition, pointing to a weakening of the first-order character. At higher Na$^+$ contents, the transition extends over a broad temperature range, reflecting an increasingly diffuse transformation driven by compositional disorder. The observed evolution of the transition temperature and estimated enthalpy can be rationalized by the substitution of Li$^+$ by the larger Na$^+$ ions, which introduces structural disorder and local lattice distortions within the Li$^+$ sublattice. This substitution effectively reduces the free-energy difference between the competing phases and disrupts the cooperative lattice rearrangements responsible for the first-order transition, thereby suppressing the transition temperature and diminishing the latent heat.

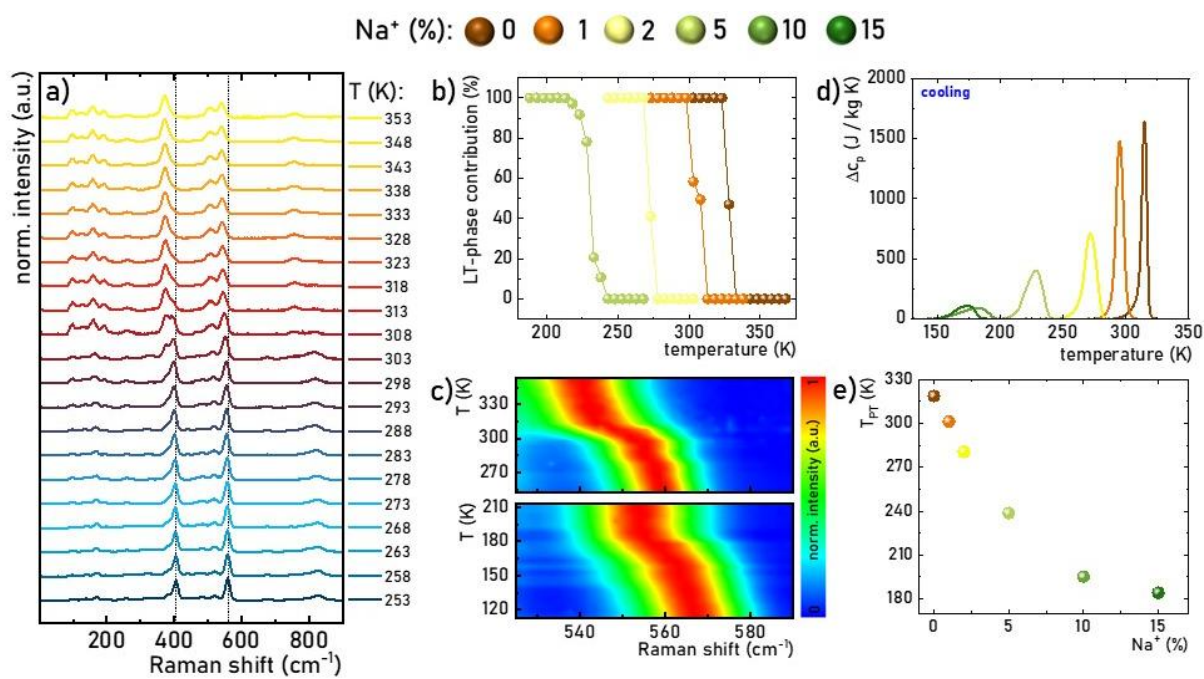

**Figure 3.** Raman spectra recorded as a function of temperature for LiYO$_2$:1%Eu$^{3+}$, 1%Na$^+$ -a). Temperature-dependent contribution of the low-temperature phase for LiYO$_2$:1%Eu$^{3+}$ samples with Na$^+$ concentrations

ranging from 0 to 5% (b). Temperature-resolved luminescence maps for LiYO$_2$:1%Eu$^{3+}$ with 1%Na$^+$ (upper panel) and 5%Na$^+$ (lower panel), focused on the 525-590 nm range -c), DSC curves measured as a function of Na$^+$ concentration - d) and phase transition temperature determined from DSC analysis as a function of Na$^+$ concentration - e).

The spectroscopic properties of Eu$^{3+}$ ions arise from intraconfigurational 4$f$-4$f$ electronic transitions. Owing to the effective shielding of the 4$f$ orbitals by the outer 5$s$ and 5$p$ shells, these transitions give rise to narrow emission lines whose spectral positions are only weakly dependent on the composition of the host lattice. Consequently, the emission spectra of Eu$^{3+}$-doped phosphors are dominated by bands corresponding to radiative depopulation of the $^5D_0$ excited state to the $^7F_J$ manifold (Figure 4a)[4,28–30]. Among these transitions, the most intense bands are typically observed near 590 nm and 620 nm, corresponding to the magnetic-dipole $^5D_0 \rightarrow {}^7F_1$ and electric-dipole $^5D_0 \rightarrow {}^7F_2$ transitions, respectively. With the exception of the $^5D_0$ and $^7F_0$ levels, all Eu$^{3+}$ energy levels undergo Stark splitting due to interactions between the 4$f$ electrons and the crystal field generated by the host lattice. The number of Stark components depends on the local symmetry of the crystallographic site occupied by Eu$^{3+}$ ions as well as on the total angular momentum quantum number $J$, generally increasing with reduction in site symmetry and increasing $J$ [4]. Importantly, the intensity of the hypersensitive electric-dipole $^5D_0 \rightarrow {}^7F_2$ transition, in contrast to the magnetic-dipole $^5D_0 \rightarrow {}^7F_1$ transition, exhibits a pronounced dependence on the local environment and typically decreases with increasing site symmetry. For this reason, Eu$^{3+}$ ions are widely employed as luminescent structural probes for assessing local symmetry and crystal-field variations in solid-state materials. In the emission spectra of LiYO$_2$:Eu$^{3+}$, in addition to the characteristic bands at approximately 585, 590, 620, 650, and 710 nm corresponding to the $^5D_0 \rightarrow {}^7F_0$-$^7F_4$ transitions, emission bands originating from the $^5D_1$ level are also observed in the 510-560 nm spectral range (Figure 4b, Figure S5-S10). Emission from the $^5D_1$ level is relatively uncommon in Eu$^{3+}$-doped materials because the small energy separation between the $^5D_1$ and $^5D_0$ levels is usually efficiently bridged by multiphonon nonradiative relaxation. However, the comparatively low phonon energy of the $LiYO_2$ host lattice suppresses this nonradiative pathway, allowing radiative $^5D_0 \rightarrow {}^7F_J$ transitions to be observed.

The strong sensitivity of $Eu^{3+}$ ions to structural changes is clearly reflected in the emission spectra of $LiYO_2:Eu^{3+}$ recorded at 83 K and 363 K, corresponding to the low-temperature (LT) and high-temperature (HT) phases of the host material (Figure 4b). The structural phase transition from the monoclinic to the tetragonal phase, accompanied by an increase in site symmetry from $C_1$ to $D_{2d}$, results in a reduction in the number of Stark components, their spectral redistribution, and a decrease in the intensity ratio of the $^5D_0 \rightarrow {}^7F_2$ to $^5D_0 \rightarrow {}^7F_1$ bands. These changes are evident for both the $^5D_1 \rightarrow {}^7F_J$ transitions (Figure 4c) and the $^5D_0 \rightarrow {}^7F_1$ (Figure 3d) and $^5D_0 \rightarrow {}^7F_2$ (Figure 4e) bands and will be discussed in detail later in this work. The small energy separation between the $^5D_1$ and $^5D_0$ levels is further manifested in the temperature dependence of the integrated emission intensities for $LiYO_2:1\%Eu^{3+}$ (Figure 4f). With increasing temperature, the emission intensity originating from the $^5D_1$ level decreases rapidly, reaching 50% of its initial value at approximately 290 K due to enhanced multiphonon relaxation. In contrast, emission from the $^5D_0$ level exhibits significantly higher thermal stability, with a 50% intensity reduction occurring only near 440 K. This behavior reflects the low probability of multiphonon depopulation from the $^5D_0$ level, arising from its large energy separation from the underlying $^7F_6$ level (~12,000 cm$^{-1}$). Moreover, thermal depopulation of the $^5D_1$ level contributes to increased population of the $^5D_0$ state, further enhancing its thermal stability. In both emission channels, a stepwise decrease in luminescence intensity is observed near 320 K, which is attributed to a phase-transition-induced shift of the $Eu^{3+}$ absorption band. Increasing the $Na^+$ concentration in $LiYO_2:1\%Eu^{3+}$, $Na^+$ leads to a progressive decrease in the luminescence intensity ratio $LIR_1$, resulting from a gradual reduction in the $^5D_0 \rightarrow {}^7F_2$ emission intensity associated with increased local site symmetry (Figure 4g):

$$LIR_1 = \frac{\int_{610nm}^{630nm} (^5D_0 \rightarrow {}^7F_2)d\lambda}{\int_{580nm}^{600nm} (^5D_0 \rightarrow {}^7F_1)d\lambda} \quad (3)$$

Interestingly, higher $Na^+$ concentrations simultaneously lead to a relative enhancement of the $^5D_1 \rightarrow {}^7F_J$ emission with respect to the $^5D_0 \rightarrow {}^7F_J$ transitions (Figure 4h).

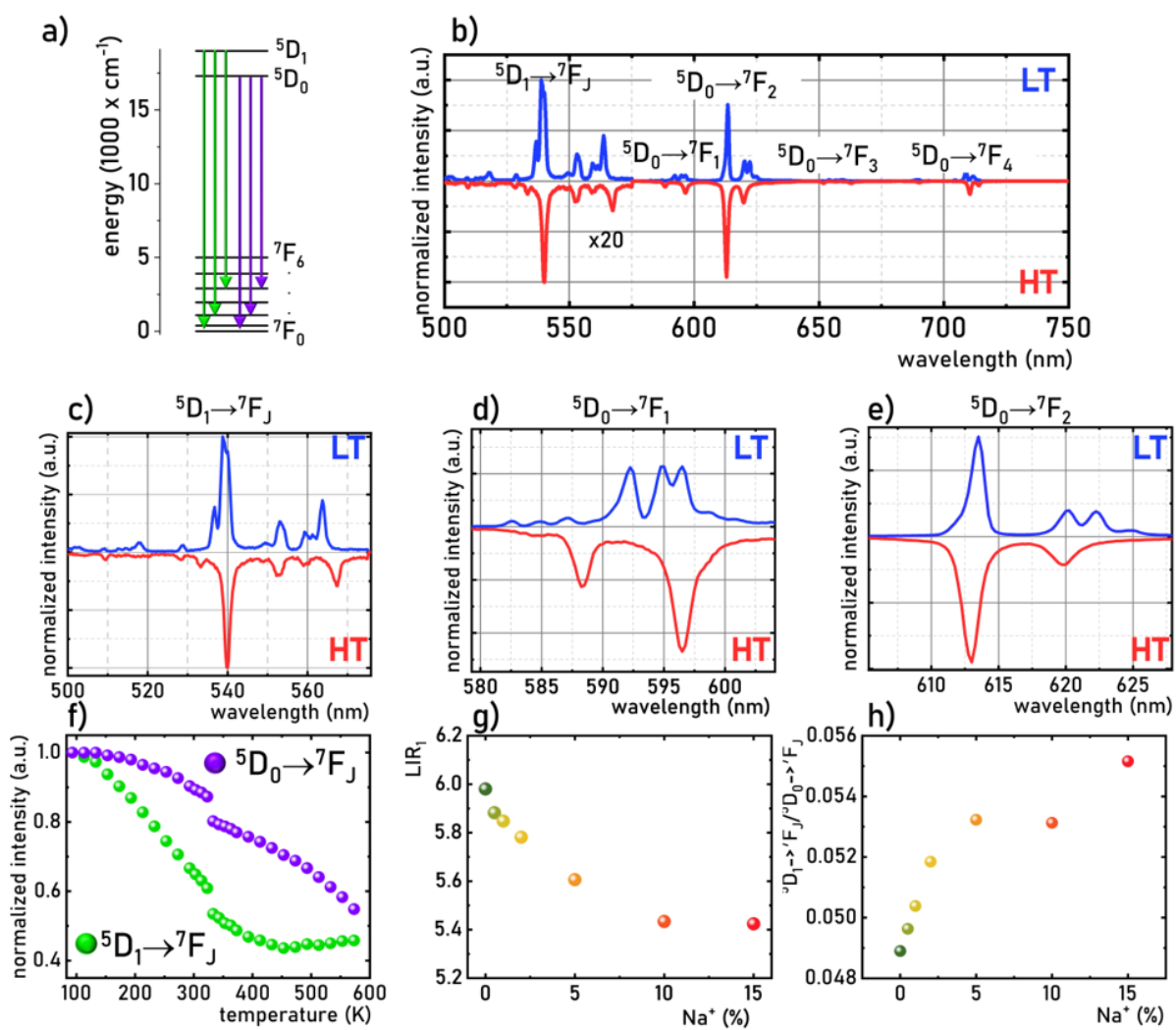

**Figure 4.** Simplified energy diagram of $Eu^{3+}$ ions in $LiYO_2:Eu^{3+}$ - a), comparison of emission spectra of $LiYO_2:Eu^{3+}$ obtained under excitation of 395 nm at 83 K (blue line) and 373 K (red line), which can be treated as a representative for LT and HT phases of $LiYO_2:1\%Eu^{3+}$ - b), and these spectra normalized in the spectral range corresponding to the emission bands associated with the $^5D_1 \rightarrow {}^7F_J$ – c), $^5D_0 \rightarrow {}^7F_1$ – d) $^5D_0 \rightarrow {}^7F_2$ – e) electronic

transitions; thermal dependence of integrated emission intensities of $^5D_1 \rightarrow {}^7F_J$ and $^5D_0 \rightarrow {}^7F_J$ emission for LiYO$_2$:1%Eu$^{3+}$ - f); the influence of Na$^+$ concentration on the *LIR$_1$* – g) and $^5D_1 \rightarrow {}^7F_J/^5D_1 \rightarrow {}^7F_J$ ratio at 93 K – h).

Changes in the spectroscopic properties of Eu$^{3+}$-doped inorganic phosphors induced by structural phase transitions have previously been exploited for ratiometric luminescence thermometry[7,11]. However, all reported systems have relied exclusively on emission from the $^5D_0$ level, resulting in thermometric operation confined to the yellow (≈590 nm) and red (≈620 nm) spectral regions. As demonstrated in the preceding analysis, LiYO$_2$:1%Eu$^{3+}$, Na$^+$ exhibits efficient luminescence originating from the $^5D_1$ level, the intensity of which increases systematically with increasing Na$^+$ concentration. This behavior enables the development of a ratiometric Eu$^{3+}$-based luminescence thermometer operating in the green spectral range, thereby substantially broadening the functional scope of phase-transition-based thermometric systems. To evaluate this concept, temperature-dependent emission spectra of LiYO$_2$:1%Eu$^{3+}$,Na$^+$ were recorded in the 500-570 nm range (Figure 5a). Representative spectra of LiYO$_2$:1%Eu$^{3+}$ reveal pronounced temperature-induced spectral shifts of the Stark components, particularly in the 560-566 nm region and around 540 nm. At temperatures corresponding to the structural phase transition, a rapid decrease in emission intensity associated with the LT phase is observed, accompanied by the emergence and growth of emission lines characteristic of the HT phase. Importantly, increasing the Na$^+$ concentration leads to a systematic reduction of the phase transition temperature from approximately 320 K for LiYO$_2$:1%Eu$^{3+}$ to about 170 K for LiYO$_2$:1%Eu$^{3+}$, 15%Na$^+$, as clearly illustrated by normalized thermal emission maps (Figure 5b-f). This shift originates from structural modifications in the local Eu$^{3+}$ environment induced by the substitution of larger Na$^+$ ions into the second coordination sphere. The temperature-dependent reshaping of the emission spectrum enables to define a ratiometric thermometric parameter, *LIR$_2$*, based on emission bands within the $^5D_1 \rightarrow {}^7F_J$ manifolds follows:

$$LIR_2 = \frac{\int_{562nm}^{565nm} \left({}^5D_1 \to {}^7F_J\right) d\lambda}{\int_{566nm}^{568nm} \left({}^5D_1 \to {}^7F_J\right) d\lambda} \tag{4}$$

The thermal evolution of $LIR_2$ exhibits a pronounced increase in the vicinity of the phase transition temperature, followed by a slight decrease at higher temperatures (Figure 5g). With increasing $Na^+$ content, the temperature at which $LIR_2$ shows the most rapid variation and reaches its maximum shifts progressively toward lower values, directly reflecting the decrease in the phase transition temperature. To quantify the observed thermal variations in $LIR_2$ the relative sensitivity $S_R$ was calculated as follows:

$$S_R = \frac{1}{LIR} \frac{\Delta LIR}{\Delta T} \cdot 100\% \tag{5}$$

In order to provide a reliable temperature readout using the luminescence thermometry it is crucial to persist monotonic change in the thermometric parameter in the thermal operating range of the thermometer. Therefore, $S_{R2}$ was evaluated within the temperature range over which $LIR_2$ varies monotonically, ensuring reliable thermometric readout. For all compositions, $S_{R2}$ exhibits a similar temperature dependence, reaching a maximum near the phase transition (Figure 5h). The highest sensitivity of 25.2% $K^{-1}$ was obtained for $LiYO_2$:1%$Eu^{3+}$ at 320 K, decreasing gradually to 4.8% $K^{-1}$ for $LiYO_2$:1%$Eu^{3+}$, 15%$Na^+$. The origin of this sensitivity reduction is discussed in detail later in the manuscript. Consistent with these trends, the temperature corresponding to maximum sensitivity ($T@S_{Rmax}$) decreases monotonically from 320 K to 148 K with increasing $Na^+$ concentration (Figure 5i).

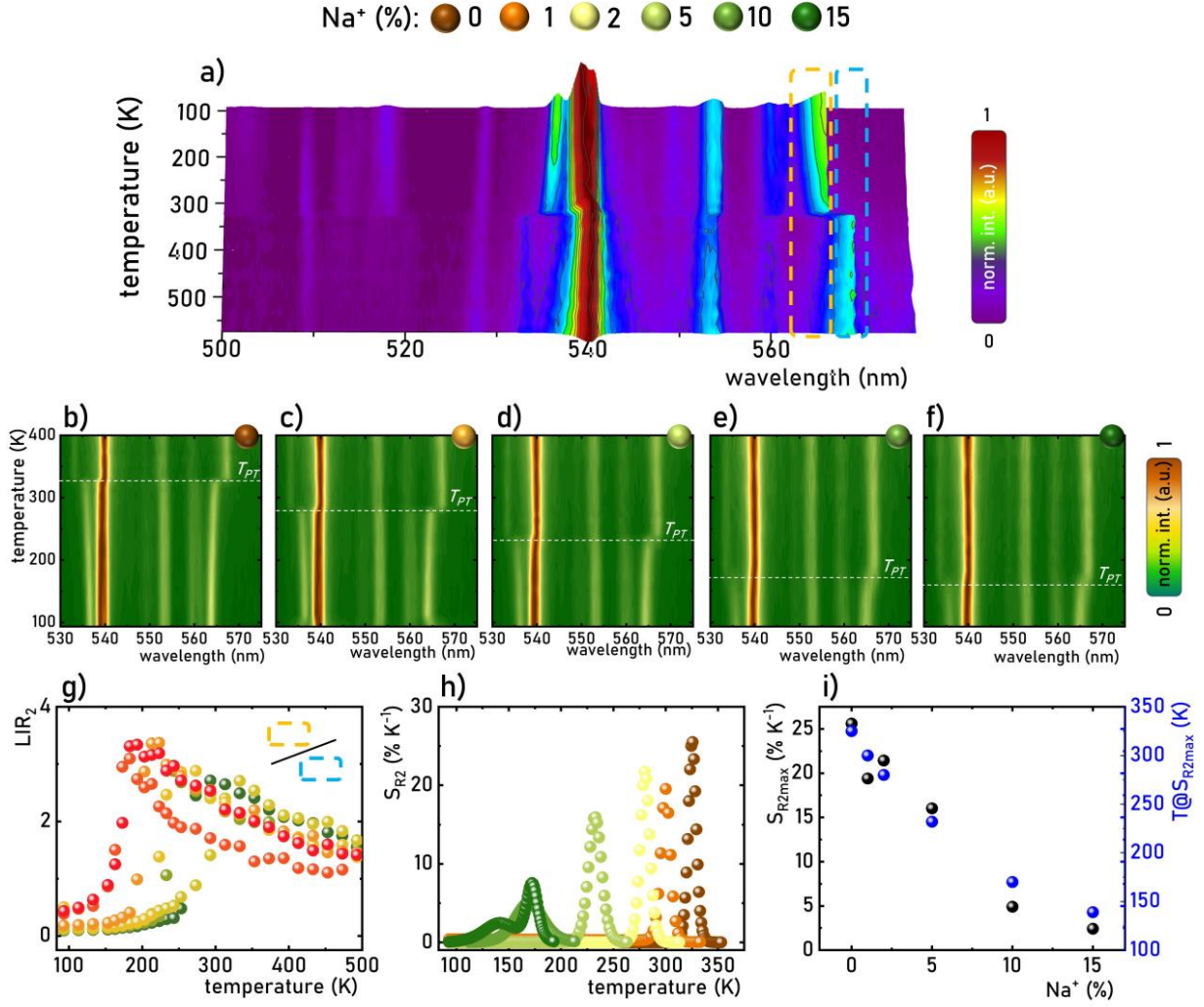

**Figure 5**. Thermal map of normalized emission spectra in the spectral range corresponding to the $^5D_1 \rightarrow ^7F_J$ transitions for LiYO$_2$:1%Eu$^{3+}$– a); the thermal maps of the normalized emission spectra of $^5D_1 \rightarrow ^7F_J$ obtained for LiYO$_2$:1%Eu$^{3+}$– b), LiYO$_2$:1%Eu$^{3+}$, 1%Na$^+$– c), LiYO$_2$:1%Eu$^{3+}$, 5%Na$^+$– d), LiYO$_2$:1%Eu$^{3+}$, 10%Na$^+$– e), LiYO$_2$:1%Eu$^{3+}$, 15%Na$^+$– f) thermal dependence of $LIR_2$ for different concentration of Na$^+$ ions – g); and corresponding $S_{R2}$ – h); the influence of Na$^+$ ions concentration on the $S_{R2max}$ and $T@S_{R2max}$ – i).

The structural phase transition in LiYO$_2$:1%Eu$^{3+}$,Na$^+$ also induces pronounced modifications in the emission bands associated with the $^5D_0 \rightarrow ^7F_1$ (Figure 6a) and $^5D_0 \rightarrow ^7F_2$ (Figure 6b) electronic transitions. Similar spectral changes are observed for the remaining $^5D_0 \rightarrow ^7F_J$ transitions; however, with increasing $J$ value, the number of Stark components increases, leading to stronger spectral overlap. This overlap complicates spectral discrimination and adversely affects the achievable $S_R$. The most pronounced thermal response is observed for the $^5D_0 \rightarrow {}^7F_1$ transition, where the number of Stark components decreases from 3 in the LT phase to 2 in the HT phase. This behavior enables the definition of the ratiometric thermometric parameter $LIR_3$ as follows:

$$LIR_3 = \frac{\int_{587nm}^{589nm} \left({}^5D_0 \rightarrow {}^7F_1\right) d\lambda}{\int_{591nm}^{592nm} \left({}^5D_0 \rightarrow {}^7F_1\right) d\lambda} \tag{6}$$

Similar to $LIR_2$, $LIR_3$ exhibits a sharp increase in the temperature range corresponding to the phase transition, followed by a slight decrease at higher temperatures (Figure 6d). Notably, the relative sensitivities obtained for $LIR_3$ exceed those derived from $LIR_2$, reaching $S_{R2max}$ = 26.2% K$^{-1}$ for LiYO$_2$:1%Eu$^{3+}$ at 320 K and decreasing to 7.6% K$^{-1}$ for LiYO$_2$:1%Eu$^{3+}$, 15%Na$^+$ at 160 K. The enhanced sensitivity of $LIR_3$ arises from the superior spectral separation of emission lines associated with the LT and HT phases. Analogous temperature-induced changes observed for the $^5D_0 \rightarrow {}^7F_2$ transition near 612 nm allow to define of $LIR_4$ (Figure 6e):

$$LIR_4 = \frac{\int_{611nm}^{612.5nm} \left({}^5D_0 \rightarrow {}^7F_2\right) d\lambda}{\int_{613nm}^{614nm} \left({}^5D_0 \rightarrow {}^7F_2\right) d\lambda} \tag{7}$$

In this case, however, substantially lower sensitivities are obtained, with $S_{R4max}$ reaching only 7.6% K$^{-1}$ for LiYO$_2$:1%Eu$^{3+}$ at 320 K (Figure 6f). Although all spectral changes originate from the same symmetry-driven mechanism, the observed differences in $S_R$ values stem from the spectral quality of the corresponding Stark components. When emission lines are closely spaced, partial spectral overlap ("intensity leakage") occurs, whereby LT-phase emission contributes to the HT-phase integration window. This effect smooths the thermal response of the luminescence intensity ratio and reduces $S_R$. To suppress intensity leakage and optimize thermometric performance, the spectral integration ranges were refined, leading to the definition of $LIR_5$, which correlates emission from both the $^5D_1$ and $^5D_0$ levels as follows:

$$LIR_5 = \frac{\int_{587nm}^{589nm} \left(^5D_0 \to {}^7F_1\right) d\lambda}{\int_{562nm}^{565nm} \left(^5D_1 \to {}^7F_2\right) d\lambda} \qquad (8)$$

Consequently, $LIR_5$ exhibits a markedly steeper thermal response near the phase transition, increasing by nearly 22-fold for LiYO$_2$:1%Eu$^{3+}$ and approximately 15-fold for LiYO$_2$:1%Eu$^{3+}$, 15%Na$^+$ (Figure 6g). This results in an exceptional maximum sensitivity of $S_{R5max} = 37.2\%$ K$^{-1}$ at 320 K for LiYO$_2$:1%Eu$^{3+}$ (Figure 6h).

Overall, these results demonstrate that LiYO$_2$:1%Eu$^{3+}$, Na$^+$ enables the realization of a ratiometric luminescent thermometer operating in three distinct spectral regions, offering flexible readout strategies with exceptionally high relative sensitivities.

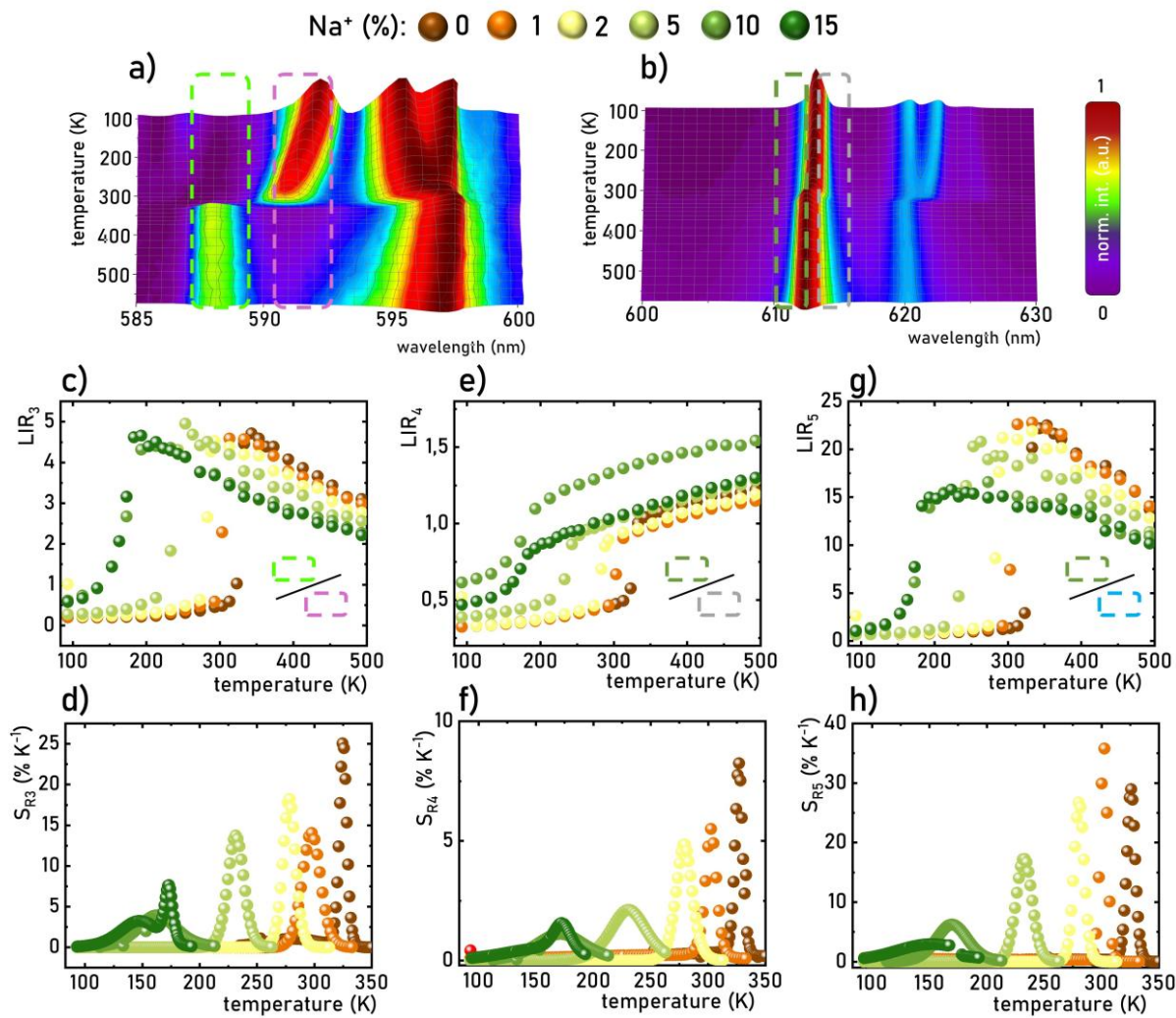

**Figure 6**. Thermal map of normalized emission spectra in the spectral range corresponding to the $^5D_0 \rightarrow ^7F_1$ – a) and $^5D_0 \rightarrow ^7F_2$ – b) transitions; thermal dependence of $LIR_3$ – c) and corresponding $S_{R3}$ – d) thermal dependence of $LIR_4$ – e) and corresponding $S_{R4}$ – f) thermal dependence of $LIR_5$ – g) and corresponding $S_{R5}$ – h) for LiYO$_2$:1%Eu$^{3+}$, Na$^+$ with different Na$^+$ ions concentration.

The exceptionally high $S_R$ observed for phase-transition-based ratiometric luminescence thermometers is highly attractive from an application standpoint. Nevertheless, a major limitation of such systems is their inherently narrow operational temperature window, which is typically restricted to the temperature range over which the structural phase transition of the host material occurs. One established strategy for shifting the phase transition temperature involves the introduction of additional dopant ions, which generate internal lattice stress and thereby modify the thermal energy required to induce the structural transformation. This approach has previously been demonstrated by substituting the host cation site occupied by Eu$^{3+}$ with ions of substantially different ionic radii. As shown in the present work, an analogous effect can be achieved through dopant incorporation in the second coordination sphere, specifically by substituting Li$^+$ ions with larger Na$^+$ ions. Importantly, this strategy is invariably accompanied by a systematic reduction in $S_R$. To elucidate the origin of this effect, a direct comparison between the temperature dependence of $S_{R5}$ and the corresponding differential scanning calorimetry (DSC) curves was performed (Figure 7a). The resulting comparison reveals a strong correspondence not only in the temperature at which the maxima occur, but also in the magnitude and full width at half maximum of both curves. This convergence clearly indicates that the thermal evolution of $S_R$ is intrinsically linked to the thermodynamic mechanism governing the structural phase transition. Owing to the substantial difference in ionic radii between Li$^+$ and Na$^+$ ions, increasing Na$^+$ content modifies the effective ionic radius at the Li$^+$ site, which can be expressed as an average radius weighted by composition:

$$R_{eff} = R_{Li} \cdot (1-x) + R_{Na} \cdot (x) \qquad (9)$$

where $R_{Li}$=0.76 Å and $R_{Na}$=1.02 Å are the ionic radii of lithium and sodium ions. As illustrated in Figure 6b, the relative increase in effective ionic radius reaches approximately 5.2% for a Na$^+$ concentration of 15%. Concurrently, the total $c_p$ change extracted from DSC measurements decreases monotonically with increasing Na$^+$ content, from approximately 120 kJ kg$^{-1}$ K$^{-1}$ for LiYO$_2$:1%Eu$^{3+}$ to about 25 kJ kg$^{-1}$ K$^{-1}$ for LiYO$_2$:1%Eu$^{3+}$, 15%Na$^+$ (Figure 7c). The progressive reduction and broadening of the calorimetric anomalies with increasing Na$^+$ content reflect a systematic weakening of the first-order character of the phase transition. As a consequence, the abrupt lattice reorganization responsible for pronounced changes in the Eu$^{3+}$ crystal-field environment becomes increasingly diffuse, which directly translates into a reduced magnitude and a broader temperature dependence of the $S_R$ response. These results highlight a fundamental trade-off between tunability of the operational temperature range and maximal sensitivity, dictated by the thermodynamics of the host lattice rather than the spectroscopic properties of the activator ion alone.

The direct correlations observed between total transition enthalpy (determined as the integral of anomalous specific heat associated with the phase transition), $T@S_{R5max}$ (Figure 7d) and $S_{R5max}$ (Figure 7e) indicate that the thermometric performance of ratiometric luminescence thermometers based on structural phase transitions in $LiYO_2$ is closely linked to the total heat associated with the transition. This finding suggests that control over the transition enthalpy provides an effective means of balancing sensitivity and operating temperature range in phase-transition-driven luminescent thermometric systems. From a thermodynamic perspective, the magnitude of the anomalous contribution to the specific heat (Δcp) associated with a structural phase transition is governed by the enthalpy change between the competing phases and the degree of cooperativity of the lattice distortion. Accordingly, the largest calorimetric anomalies are expected for systems exhibiting well-defined first-order transitions characterized by long-range correlated structural rearrangements. In $LiYO_2$, this condition is fulfilled when the $Li^+$ sublattice remains structurally homogeneous, enabling collective lattice distortions. Partial substitution of $Li^+$ by the larger $Na^+$ ions provides an effective route for tuning the transition temperature; however, it typically introduces static compositional disorder and local lattice strain. While even small $Na^+$ concentrations significantly suppress the transition temperature, increasing dopant content progressively reduces the latent heat and broadens the specific heat anomaly, reflecting a weakening of the first-order character of the transition. This behavior indicates that dopant-induced disorder lowers the free-energy difference between the low- and high-temperature phases and disrupts the cooperative nature of the structural transformation. Consequently, maximizing Δcp in phase-transition-based materials requires minimizing quenched disorder within the sublattice directly involved in the transition, while maintaining a sufficiently large free-energy barrier between the competing phases. Excessive substitutional disorder, although beneficial for extending the accessible temperature range, inherently limits the magnitude of the excess specific heat, resulting in a trade-off between transition sharpness and operational temperature window that can be systematically controlled through dopant concentration.

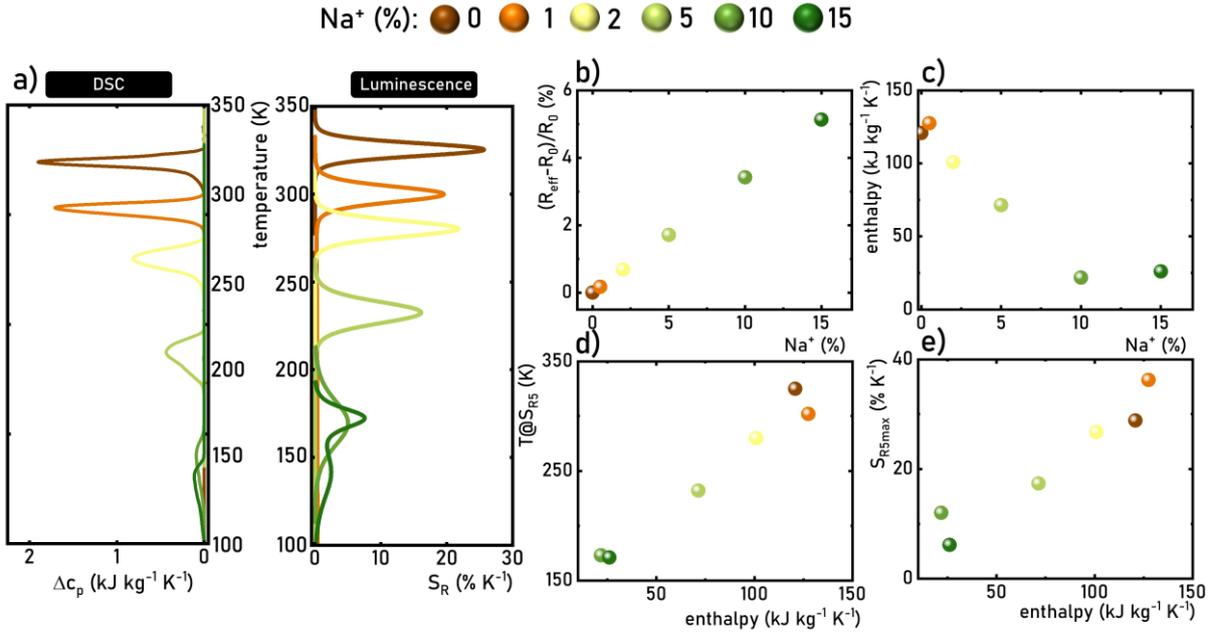

**Figure 7**. The comparison of the thermal dependence of $S_{R5}$ and $\Delta c_p$ for different concentrations of Na$^+$ ions in LiYO$_2$:Eu$^{3+}$, Na$^+$ – a); the influence of Na$^+$ ions concentration on the $(R_{eff}-R_0)/R_0$ – b); and $\Delta c_p$ – c); the $T@S_{R5max}$ – d); and $S_{Rmax}$ – e) as a function of $\Delta c_p$ for LiYO2:1%Eu$^{3+}$, Na$^+$.

**Conclusions**

In this study, the spectroscopic properties of LiYO$_2$:1%Eu$^{3+}$, Na$^+$ were systematically investigated as a function of temperature and Na$^+$ concentration in order to evaluate the feasibility of tuning the thermometric performance of phase-transition-based luminescence thermometers through substitution in the second coordination sphere of Eu$^{3+}$ ions. LiYO$_2$:1%Eu$^{3+}$, Na$^+$ undergoes a first-order structural phase transition from a low-temperature monoclinic phase to a high-temperature tetragonal phase, with the transition temperature decreasing from 320 K for LiYO$_2$:1%Eu$^{3+}$ to approximately 160 K for LiYO$_2$:1%Eu$^{3+}$, 15%Na$^+$. This shift is attributed to structural modifications arising from the substitution of Li$^+$ by larger Na$^+$ ions, which shorten the average Eu$^{3+}$-O$^{2-}$ distance and destabilize the low-temperature phase.

The phase transition induces pronounced changes in the spectroscopic behavior of Eu$^{3+}$ ions, manifested by a reduction in the number of Stark components and spectral redistribution of emission bands associated with radiative depopulation of the $^5D_1$ and $^5D_0$ levels. Opposite monotonic thermal variations in emission intensities originating from the low- and high-temperature phases enable the construction of ratiometric luminescence thermometers. Notably, this work demonstrates for the first time the use of the $^5D_1 \rightarrow {}^7F_J$ transitions for phase-transition-based thermometry, enabling operation in the green spectral range. The maximum relative sensitivity reaches 25.2% K$^{-1}$ for LiYO$_2$:1%Eu$^{3+}$ at 320 K and decreases systematically with increasing Na$^+$ content. Additional thermometric modes operating in the yellow and red spectral regions, based on the $^5D_0 \rightarrow {}^7F_1$ and $^5D_0 \rightarrow {}^7F_2$ transitions, yield maximum sensitivities of 26.2% K$^{-1}$ and 7.6% K$^{-1}$, respectively. Optimization of spectral integration ranges to suppress intensity leakage enables further enhancement, yielding $S_{R5max}$ = 37.2% K$^{-1}$ for LiYO$_2$:1%Eu$^{3+}$ at 320 K. For all thermometric modes, increasing Na$^+$ concentration results in a concurrent reduction of the phase transition temperature and $S_{Rmax}$. This effect is directly linked to a modification of the phase transition mechanism, as evidenced by calorimetric analysis. The observed correlations between $S_{Rmax}$, $T@S_{Rmax}$, and total transition enthalpy indicate that maximizing sensitivity favors host materials with high heat capacity. Importantly, modulation of thermometric performance via second-sphere substitution provides a more efficient and cost-effective route than direct lanthanide-site substitution, as significantly larger ionic-radius mismatches can be achieved at lower dopant concentrations. These findings demonstrate that precise control of host lattice composition in the second coordination sphere offers a powerful strategy for engineering high-performance phase-transition-based luminescence thermometers and suggest that, in some cases, identifying new host materials with inherently suitable phase transition temperatures may be more advantageous than extensive dopant-driven tuning.

**Acknowledgements**

This work was supported by the National Science Center (NCN) Poland under project no. DEC-UMO-2022/45/B/ST5/01629. M.Sz. gratefully acknowledges the support of the Foundation for Polish Science through the START program.